\documentclass[lettersize,journal]{IEEEtran}
\usepackage{amsmath,amsfonts}

\usepackage[T1]{fontenc}
\usepackage{array}
\usepackage[caption=false,font=normalsize,labelfont=sf,textfont=sf]{subfig}
\usepackage{textcomp}
\usepackage{stfloats}
\usepackage{url}
\usepackage{verbatim}
\usepackage{graphicx}
\usepackage{cite}
\usepackage{epstopdf}
\usepackage{amsmath,amsthm, amssymb}
\usepackage{bm}
\usepackage{float}
\usepackage{lineno}
\usepackage{setspace} 
\usepackage{stfloats}
\usepackage{amsmath}
\allowdisplaybreaks[0]
\makeatletter
\newif\if@restonecol
\makeatother

\usepackage[linesnumbered,ruled,vlined]{algorithm2e}
\usepackage{algpseudocode}
\usepackage{amsmath}

\usepackage{fancyhdr}

\begin{document}
\onecolumn 

\title{Vehicle Selection for C-V2X Mode 4 Based Federated Edge Learning Systems}

\author{Qiong Wu,~\IEEEmembership{Senior Member,~IEEE}, Xiaobo Wang, Pingyi Fan,~\IEEEmembership{Senior Member,~IEEE}, Qiang Fan, \\  Huiling Zhu,~\IEEEmembership{Senior Member,~IEEE}, Jiangzhou Wang,~\IEEEmembership{Fellow,~IEEE}
\thanks{

Qiong Wu and Xiaobo Wang are with the School of Internet of Things Engineering, Jiangnan University, Wuxi 214122, China, and also with the State Key Laboratory of Integrated Services Networks (Xidian University), Xi'an 710071, China (e-mail: qiongwu@jiangnan.edu.cn, xiaobowang@stu.jiangnan.edu.cn).

Pingyi Fan is with the Department of Electronic Engineering, Beijing National Research Center for Information Science and Technology, Tsinghua University, Beijing 100084, China (Email: fpy@tsinghua.edu.cn).

Qiang Fan is with Qualcomm, San Jose, CA 95110, USA (e-mail: qf9898@gmail.com).

Huiling Zhu and Jiangzhou Wang are with the School of Engineering, University of Kent, CT2 7NT Canterbury, U.K. (Email: H.Zhu@kent.ac.uk, j.z.wang@kent.ac.uk).

}
\thanks{}
\thanks{}}

\markboth{}
{WU \MakeLowercase{\textit{et al.}}: Vehicle Selection for C-V2X Mode 4 Based Federated Edge Learning Systems}


\maketitle

\begin{abstract}
Federated learning (FL) is a promising technology for vehicular networks to protect vehicles' privacy in Internet of Vehicles (IoV). Vehicles with limited computation capacity may face a large computational burden associated with FL. Federated edge learning (FEEL) systems are introduced to solve such a problem. In FEEL systems, vehicles adopt the cellular-vehicle to everything (C-V2X) mode 4 to upload encrypted data to road side units' (RSUs)' cache queue. Then RSUs train the data transmitted by vehicles, update the locally model hyperparameters and send back results to vehicles, thus vehicles’ computational burden can be released. However, each RSU has limited cache queue. To maintain the stability of cache queue and maximize the accuracy of model, it is essential to select appropriate vehicles to upload data. The vehicle selection method for FEEL systems faces challenges due to the random departure of data from the cache queue caused by the stochastic channel and the different system status of vehicles, such as remaining data amount, transmission delay, packet collision probability and survival ability. This paper proposes a vehicle selection method for FEEL systems that aims to maximize the accuracy of model while keeping the cache queue stable. Extensive simulation experiments demonstrate that our proposed method outperforms other baseline selection methods.
\end{abstract}


\section{Introduction}
\IEEEPARstart{W}{ith} the development of the information and communication technology of Internet of Vehicles (IoV), vehicles use on-board sensors to collect data and process them to achieve massive interaction information \cite{1, 2, 3, 4}, which can satisfy the requirements of different vehicular applications, such as intelligent transportation and object identification \cite{5, 6}. However, vehicles are often unwilling to send their data to other vehicles, as vehicle data are private, which makes it difficult to fully meet the needs of vehicle applications. Federated learning (FL) can protect vehicle privacy by sending locally trained models instead of raw data among vehicles to solve such problem \cite{7, 8}. In particular, each vehicle trains a local model based on its own data, and then the local model is uploaded to the central cloud. After that, the central cloud aggregates the received local models. Therefore, the global model is obtained and can then be sent to the vehicles \cite{9}. By doing this, an accurate model can be achieved and vehicles' data can be preserved. The functions for the federated learning can be various, e.g., autonomous driving, location awareness, mobile biometrics, mobile augmented reality, and analysis of sensing signal \cite{10, 11, 12}.

Vehicles have intensive computing tasks to process, which would consume lots of computation resources \cite{13}. However, the computational resources of a vehicle depend on the limited energy resources such as gasoline or electric power of the vehicle, thus the computation capability of vehicles are limited. Therefore, a high computational burden would be caused on the vehicles in FL due to vehicles' limited computation capability \cite{14, 15, 16}. 
Federated edge learning (FEEL) systems are introduced to solve such a problem.

The FEEL system is a three-tier infrastructure that includes a central cloud, road side units (RSUs) and vehicles. Each RSU is connected with an edge server and has a cache queue. Each vehicle is equipped with a shuffler. Note that in the FEEL system, the RSUs would not harm the users' privacy by equipping a shuffler in the vehicles, which can encrypt the raw data uploaded by the vehicles in a uniformly random permutation. 
In the FEEL system, each vehicle uploads the encrypted data to the RSUs, then the RSUs and central cloud adopt the federated learning to train an accurate model while protecting the privacy of vehicles. Specifically, the RSUs train the local model based on the uploaded data. After that, all RSUs send the locally trained model hyperparameters to the central cloud, which aggregates them to get the global model. Then, the central cloud broadcasts the global model to the RSUs for further training. For the FEEL system, the vehicles with limited computation capability are only responsible for uploading data, while the RSUs and central cloud with sufficient computation capability apply the federated learning. Hence, FEEL system can reduce the computational burden on the vehicles while protecting privacy \cite{17, 18}. 

Each vehicle adopts the cellular-vehicle to everything (C-V2X) mode 4, which is the key enabling technology for intelligent mobility and autonomous driving \cite{19, 20}, to communicate with the RSU. 
Note that we consider the scenarios that have high requirements for communication range, stability, and latency to satisfy applications with high safety requirements such as autonomous driving, hence we choose C-V2X mode 4 in this paper.

Each RSU's cache queue is limited. The cache queue can become overloaded and unstable if many vehicles upload data. On the other hand, if few vehicles upload data, the cache queue would have insufficient data for training and the accuracy of the model will suffer. In order to maintain queue stability and obtain accurate models, we have to fill the RSU's cache queue with vehicle data by taking into account the queue capacity and selecting appropriate vehicles to upload data. In the FEEL system, the channel condition is stochastic and RSU sends back results to vehicles only when the channel condition is sufficiently good, thus the data depart the cache queue randomly. Moreover, different vehicles may have different system status including remaining data amount, transmission delay, packet collision probability and survival ability (i.e., duration of staying in a RSU's communication range), where the packet collision probability is the important metric of C-V2X mode 4. As the status of each vehicle includes several entities, it is difficult to select the appropriate number of vehicles for data uploading considering the random departure of data and different system status. To the best of our knowledge, there are no related works considering the use of FEEL for vehicular scenarios and no related works to improve the accuracy of the model by considering different system status and queue stability for vehicle selection in the C-V2X mode 4 based FEEL system, which motivated us to undertake this work. In fact, compared with the traditional communication methods, C-V2X has the advantages including wider coverage, higher reliability, lower latency, scalability and lower cost.


In this paper, considering the different system status of vehicles and random departure of data, we propose a vehicle selection method in the C-V2X mode 4 based FEEL system, which can regulate the stability of the cache queue and improve the accuracy of the model\footnote{The source code has been released at: https://github.com/qiongwu86/Vehicle-selection-for-C-V2X.git}. The main contributions are summarized as follows.

\begin{itemize}
\item[1)]We consider a C-V2X mode 4 based FEEL system which can protect vehicles' privacy and relieve the vehicles' computational burden, and propose a vehicle selection method that can maximize the accuracy of the model and maintain the stability of cache queue in the C-V2X mode 4 based FEEL system. In the proposed method, multiple entities representing the status of a vehicle are jointly considered, which includes remaining data amount, transmission delay, packet collision probability and survival ability of the vehicle.

\item[2)]For each RSU, we consider the random departure of data and determine the optimal number of selected vehicles to ensure the stable cache queue according to Lyapunov's control theorem.

\item[3)]Extensive simulation experiments are conducted to demonstrate that our proposed method outperforms other baseline methods.

\end{itemize}

The rest of this paper is organized as follows. In Section~\ref{Related Work} we first review the related work and then describe the system model in Section~\ref{System Model}. The problem is formulated in Section~\ref{Problem Formulation}. The vehicle selection method is proposed in Section~\ref{Vehicle Selection Method}. Section~\ref{Simulation Results and Analysis} shows the simulation results and Section~\ref{Conclusion} concludes this paper.

\begin{figure*}[htbp]
\centering
\includegraphics[width=12cm, scale=1.00]{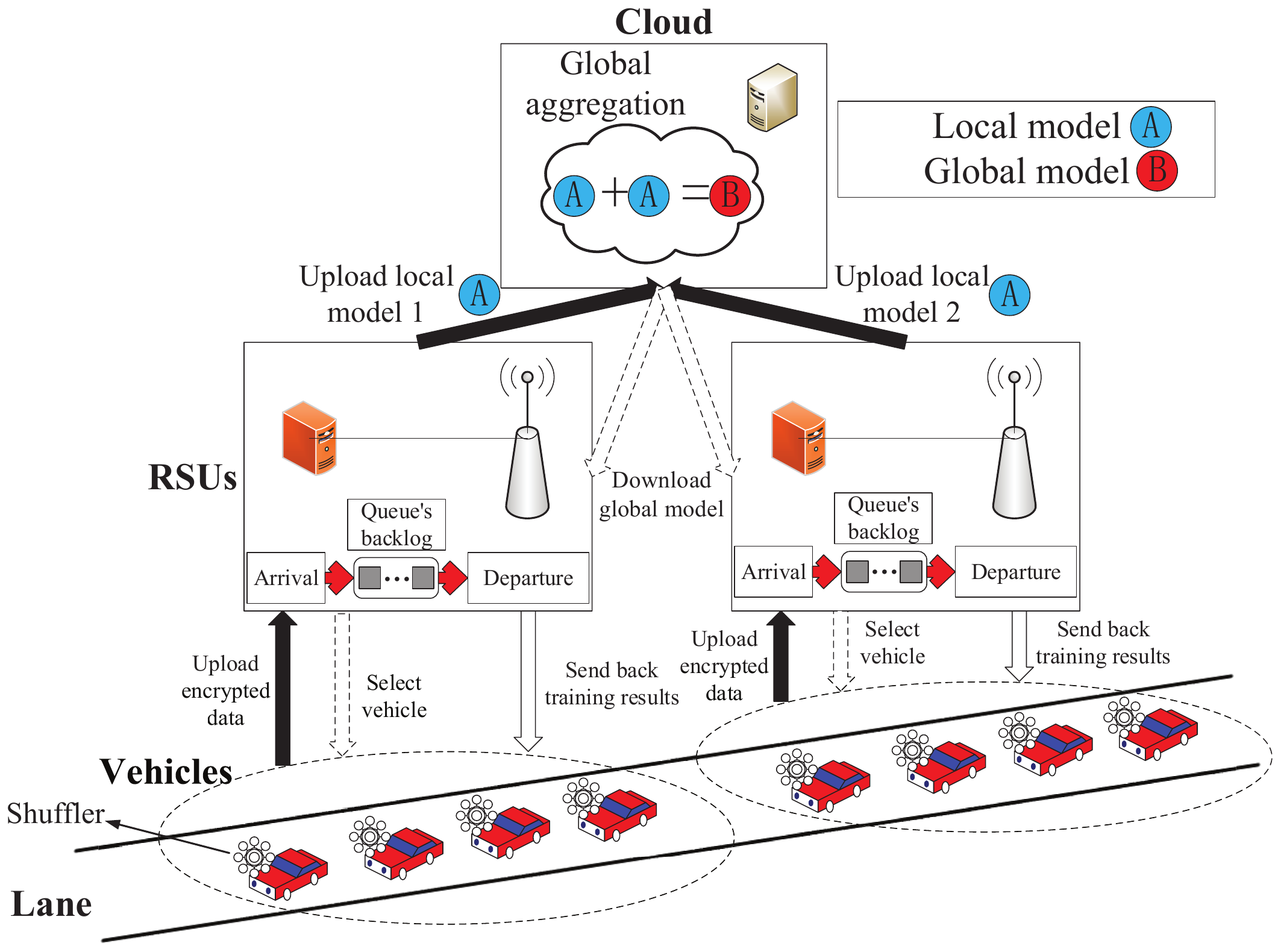}
\caption{System model.}
\label{fig1}
\vspace{-0.7cm}
\end{figure*}

\section{Related Work}
\label{Related Work}
In this section, we first review the works related to vehicle selection in the FL framework, and then the works related to vehicle selection in FEEL systems.

Most of the related works considered vehicle selection in the FL. In \cite{21}, Xiao \textit{et al.} considered the vehicle position and velocity to formulate a min-max optimization problem which jointly optimizes the onboard computation capability, transmission power and local accuracy of the model, and proposed a greedy algorithm to select vehicles with higher image quality dynamically to achieve the minimum cost in the worst case of FL. In \cite{22}, Li \textit{et al.} addressed the problem of the massive map data transmission and strict demand for privacy to propose a joint optimization scheme of participant selection and resource allocation for FL, which minimizes long-term training delay with limited energy consumption. In \cite{23}, Fu \textit{et al.} considered the fast and accurate response of a connected autonomous vehicle (CAV) in the complex traffic environment and proposed a selective federated reinforcement learning (SFRL) strategy to achieve online knowledge aggregation strategy to improve the accuracy and environmental adaptability of the autonomous driving model. In \cite{24}, Huang \textit{et al.} considered the clients' availability, unknown and stochastic training time, as well as the dynamic communication status, where clients include tablet PCs, vehicles, phones, WiFis and proposed a combinatorial multi arm bandit (C$^{2}$MAB)-based method to estimate the model exchange time between each client and the server, then they designed a reputation based client selection with fairness (RBCS-F) algorithm to model the fairness guaranteed client selection. In \cite{25}, Lu \textit{et al.} proposed an asynchronous FL scheme by adopting deep reinforcement learning (DRL) for vehicle selection to enhance the security and reliability of model parameters. In \cite{26}, Saputra \textit{et al.} considered the challenges including dynamic activities and diverse quality-of-information (QoI) from a large number of smart vehicles (SVs), vehicular service provider (VSP)'s limited payment budget and profit competition among SVs to propose a SV selection method based on dynamic FL for an IoV network to address these challenges. 
In \cite{27}, Cassará \textit{et al.} considered the problems including the limited computation and communication workloads caused by extracting the massive amount of multi-modal data generated by autonomous vehicles (AVs) and proposed a federated feature selection (FFS) algorithm to solve these problems. 
In \cite{28}, AbdulRahman \textit{et al.} considered the heterogeneity of the client devices in an internet-of-things (IoT) environment and their limited communication, computation resources of the devices, where the client devices are various IoT devices, and proposed a multicriteria-based approach for client selection in FL. In \cite{29}, Kang \textit{et al.} considered the incentive mechanisms for participating in training and designed a reputation-based worker (i.e., vehicles or mobile phones.) selection scheme for reliable FL by using a multiweight subjective logic model. In \cite{30}, Zhou \textit{et al.} proposed a two-layer FL model and designed a novel multi-layer heterogeneous vehicles' model selection and aggregation scheme to achieve a more efficient and accurate learning while ensuring data privacy protection and reduce communication overheads. In \cite{31}, Liu \textit{et al.} considered the intermittent connectivity for the distributed vehicles in the fifth generation (5G) and beyond 5G (B5G) vehicular network and proposed a vehicle selection scheme in the uplink-downlink decoupled 5G/B5G networks to enhance energy efficiency. However, the above works select vehicles based on FL, which trains local models by vehicles, thus imposing a large computational burden for vehicles.

Few related works considered vehicle selection in FEEL systems. In \cite{32}, we considered the status of vehicles including remaining data amount, communication quality, remaining energy and survival ability in the FEEL system, and proposed a vehicle selection scheme to maximize the accuracy of the model and maintain the stable cache queue. However, this work has not considered C-V2X mode 4 technology, which is the key enabling communication technology for intelligent transportation and autonomous driving. Compared with the existing work, this is the first work considering C-V2X mode 4 technology in FEEL system to select vehicle.

\section{System Model}
\label{System Model}

\subsection{System scenario}

Fig.~\ref{fig1} depicts a C-V2X mode 4 based FEEL system. The system comprises vehicles, RSUs and a central cloud, where the central cloud covers multiple RSUs. The RSUs are deployed along the road, and each RSU is connected with an edge server. The communication range of each RSU is a circle of radius $R$, and there are $K$ vehicles in it driving in the same direction. Each vehicle is equipped with a shuffler to encrypt its raw data in a uniformly random permutation before uploading, thus the RSUs would not harm vehicles' privacy. Each RSU has a cache queue with the maximum length of $Q_{max}$ to store data, where the cache policy of each queue is the first-come, first-served (FCFS). When the data in the queue surpass the queue's maximum length, they will be discarded.

The duration of a vehicle driving in the communication range of the RSUs is split into equal time intervals. In each time interval, each vehicle in a RSU first calculates the system status including the remaining data amount, transmission delay, packet collision probability and survival ability, then sends the system status to RSU. Afterwards, the RSU selects some vehicles to upload data. In particular, the RSU first determines the number of selected vehicles to avoid overflowing and then identifies which vehicles are selected based on their system status. After vehicles are selected, the selected vehicles will adopt the shuffler to encrypt their raw data in a uniformly random permutation, and forward the encrypted data to the RSU. The forwarded data are stored in the RSU's cache queue and for the RSU to train the model locally. At the end of the training, the RSU updates the local model. Next, the RSU feedbacks the training results to the corresponding vehicles and discards the training data simultaneously, if the conditions of the channel between the RSU and the selected vehicles are sufficiently good. Otherwise, the data stay in the queue until the conditions of the channel between the RSU and the selected vehicles are sufficiently good. Up to now, the process in the time interval is finished. The above process is repeated until all vehicles in the RSU communication range have no data for uploading. After that, the updated local model at the RSU is uploaded to the central cloud. After the central cloud collects the local models from all RSUs, the local models would be aggregated to update the global model. Finally, the RSUs download the updated global model from the central cloud for the next training. During the above process, the communication between vehicles and RSUs employs C-V2X mode 4.

\subsection{C-V2X Mode 4}

\begin{figure}[htbp]
\centering
\includegraphics[width=12cm, scale=1.00]{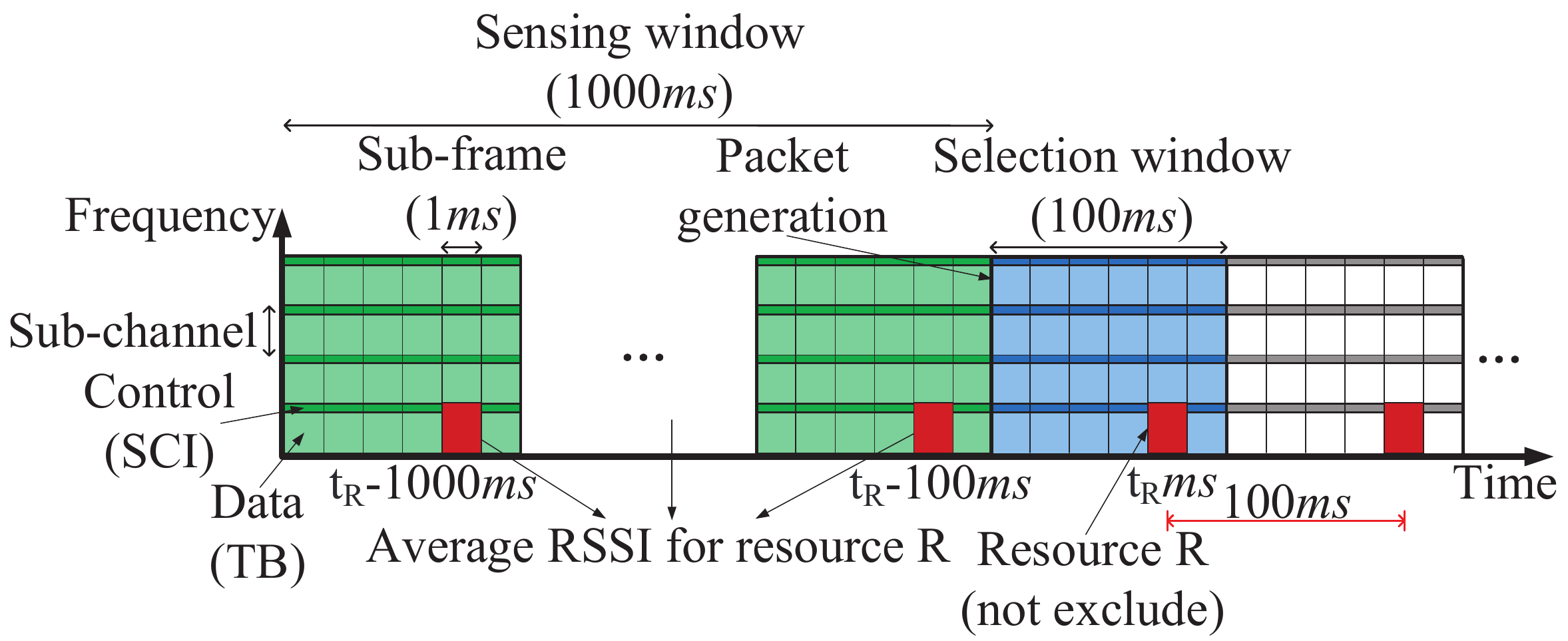}
\caption{SPS process.}
\label{fig2}
\end{figure}

The vehicle adopts the sensing-based semi-persistent scheduling (SPS) scheme of the C-V2X mode 4 to reserve and select resources \cite{33, 34, 35}. The process of SPS is performed in a time-frequency domain, where the sub-channels and sub-frames are divided into the frequency domain and time domain, respectively. The duration of each sub-frame is 1 ms. A single sub-frame resource consists of a sub-channel and a sub-frame. The single sub-frame resource is referred to as resource for brief. Each resource is adopted to transmit a data packet and the corresponding control information, where a packet is transmitted through a transport block (TB) and the control information is transmitted through a sidelink control information (SCI) \cite{36}. Fig.~\ref{fig2} shows the SPS process of C-V2X mode 4 when the number of sub-channels is 4, where each block is the resource.

The process of the SPS scheme comprises two phases, i.e., sensing and selection, which are introduced as follows.

\subsubsection{Sensing}
The vehicle monitors other vehicles' SCI in a sensing window to sense the resources which are occupied or will be occupied by other vehicles, and marks them as busy. Moreover, the vehicle receives the reference signal received powers (RSRPs) and received signal strength indicators (RSSIs) of the occupied resources from other vehicles. Next, the vehicle will reserve and select resources based on the sensed resources as well as the received RSRPs and RSSIs.

\subsubsection{Selection}The vehicle adopts the following three steps to reserve and select resources.

\noindent \textbf{Step 1:} The vehicle first reserves all resources within the selection window, which is a resource reservation interval (RRI), denoted as $\Gamma$ ms, after a packet is generated.

\noindent \textbf{Step 2:} The vehicle creates a list $L_A$ to reserve the available resources, which includes all resources in the selection window excluding the following three kinds of resources:

\noindent \textbf{a.} The corresponding resources used by the vehicle during the sensing window should be excluded.

\noindent \textbf{b.} The corresponding resources that have been occupied or will be occupied by other vehicles in the sensing window should be excluded.

\noindent \textbf{c.} If the average RSRP of the resources for a sub-channel is higher than a sensing power threshold specified by the 3GPP \cite{37}, the resources for the sub-channel should be excluded.

When Step 2 is finished, the condition that $L_A$ contains at least 20$\%$ of all resources in the selection window should be satisfied. Otherwise, the RSRP threshold is increased by 3 dB and Step 2 is iteratively executed until the condition is satisfied.

\noindent \textbf{Step 3:} The vehicle creates a list $L_C$ to reserve the candidate resources, where the size of $L_C$ equals 20$\%$ of all the resources in the selection window. The average RSSI of the resources in $L_A$ at $t_R-\Gamma$, $t_R-2\Gamma$, ...,  $t_R-1000$ is calculated, respectively, where $t_R$ is the time to transmit the packet. The resources are selected from the lowest average RSSI to the highest average RSSI in turn as the resources of $L_C$ until the size of $L_C$ is reached. After that, the vehicle randomly selects one of the candidate resources in $L_C$. Then the vehicle selects a random value $RC$ from $[R_l, R_h]$ as the value of the resource counter, where $R_h$ and $R_l$ are upper and lower bounds of resource counter and depend on the packet transmission frequency $f$ $(f=1000/\Gamma)$. Then the vehicle reserves the selected resource and the resources after each RRI until the number of the reserved resources reaches $RC$. Then the vehicle adopts this selected resource to transmit the packet and the value of the resource counter is decreased by one. After each RRI, if the vehicle has a packet to transmit, it will adopt the corresponding reserved resource to transmit the packet. After each transmission, the value of the resource counter is decreased by one. When the value of the resource counter equals zero, the vehicle would repeat the SPS process to select and reserve new resources for transmission \cite{19}.

\section{Problem Formulation}
\label{Problem Formulation}
The Lyapunov control theorem is employed in this section to determine the optimal number of selected vehicles according to the queue's backlog.

\begin{figure}[htbp]
\centering
\includegraphics[width=12cm, scale=1.00]{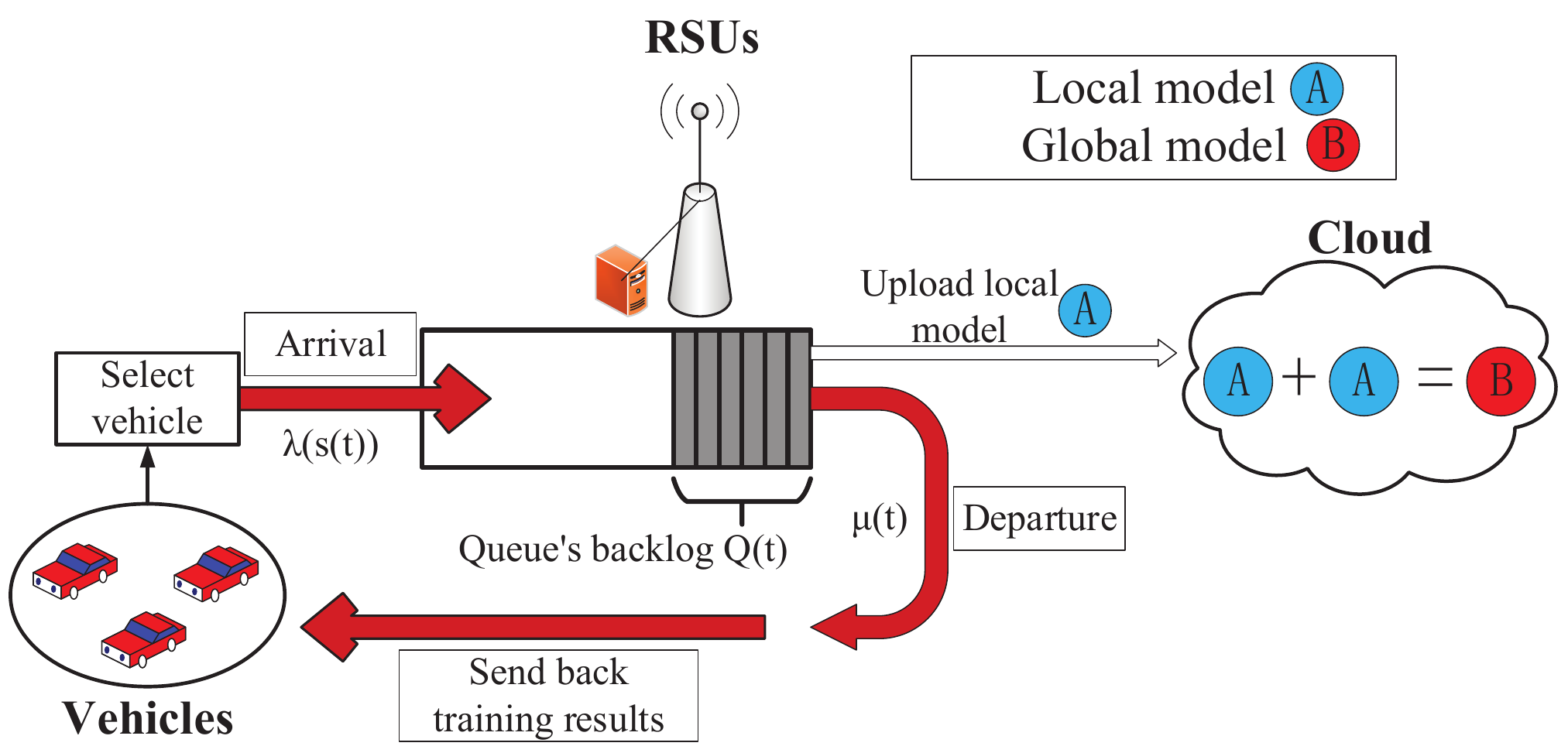}
\caption{Cache queue of RSU.}
\label{fig3}
\end{figure}

According to the queueing theory, the data arrival and departure in the current time interval determines the queue's backlog in the next time interval, thus the RSU's cache queue is expressed as
\begin{equation}
Q(t+1) = \max \{Q(t)+\lambda(t)-\mu(t),0\},
\label{eq1}
\end{equation}
where $\lambda(t)$ is the arrival data amount at RSU in the current time interval $t$, $\mu(t)$ is the departure data amount at RSU in the current time interval $t$, $Q(t)$ is the queue's backlog in the current time interval $t$. The RSU's cache queue is shown in Fig.~\ref{fig3}.

Considering each vehicle uploads the same data amount, the number of selected vehicles in time interval $t$, which is denoted as $s(t)$, determines the arrival data amount at the RSU. When the condition of the channel is sufficiently good, the RSU sends back the results to the vehicles. Moreover, the condition of the channel is random in each time interval, thus $\mu(t)$ is random. Denote $\lambda(s(t))$ as the arrival data amount at the RSU under $s(t)$, which is expressed as
\begin{equation}
\lambda(s(t))=D_a \times s(t),
\label{eq2}
\end{equation}
where $D_a$ is the uploaded data amount from each vehicle in each time interval.

The objective for each RSU is maximizing the accuracy of the model, which means selecting as many vehicles as possible to upload data, and guaranteeing that the queue's backlog of each RSU would not exceed the queue's maximum length. Let $U(s(t))$ be the expected accuracy's utility function under $s(t)$, $\gamma$ be a parameter to trade off the utility function and Lyapunov drift, $s^*(t)$ be the optimal number of selected vehicles in time interval $t$, $\boldsymbol{X}$ be the set of the selected vehicles' number ($\boldsymbol{X}=\{0,1,2,\cdots,K\}$), $T$ be the maximum number of time intervals and $Q_{max}$ be the queue's maximum length. According to the theorem of Lyapunov control \cite{38}, the optimization objective can be expressed as
\begin{equation}
\begin{aligned}
&s^*(t)\leftarrow\\ &\mathop{\arg\max}\limits_{s(t)\in \boldsymbol{X}}\{\gamma\cdot U(s(t))+Q(t)\cdot\left[\lambda(s(t))-\mu(t)\right]\},
\end{aligned}
\label{eq3}
\end{equation}
\begin{equation}
s.t.\quad \lim\limits_{T\rightarrow \infty}\frac{1}{T}\sum_{t=0}^{T-1}Q(t)\leq Q_{max},
\label{eq4}
\end{equation}
where $Q(t)\cdot\left[\lambda(s(t))-\mu(t)\right]$ is the Lyapunov drift \cite{38} and Eq.~\eqref{eq4} is the queue stability constraint.

Based on the above problem formulation, $s^*(t)$ can be determined by the theorem of Lyapunov control base on Eqs.~\eqref{eq1}-\eqref{eq4}. The problem aims to find $s^*(t)$ to satisfy Eq.~\eqref{eq3} under the queue stability constraint Eq.~\eqref{eq4}.

\section{Vehicle Selection Method}
\label{Vehicle Selection Method}
In this section, we present a method for selecting vehicles. In particular, each vehicle in time interval $t$ in the communication range of the RSU calculates its system status consisting of the remaining data amount, transmission delay, packet collision probability and survival ability, i.e., $A^k(t)$, $D^k(t)$, $P^k_{col}(t)$ and $S^k(t)$, then sends the system status to the RSU. Afterwards, the RSU selects vehicles based on its queue's backlog and vehicles' different system status. Next, we first describe the calculations of the system status and then present the proposed vehicle selection method.

\subsection{System status}
A time interval is sufficiently long for each vehicle to calculate the system status.

\subsubsection{Remaining data amount}
\
\newline
\indent
The remaining data amount of vehicle $k$ in each time interval $t$ represents the data amount carried by vehicle $k$ in each time interval $t$, which is denoted as $A^k(t)$. If vehicle $k$ is selected to transmit data in time interval $t$, the data amount carried by vehicle $k$ in previous time interval $t-1$ is decreased by the transmitted data amount $D_a$, thus $A^k(t)$ is calculated as
\begin{equation}
A^k(t) = A^k(t-1) - D_a.
\label{eq5}
\end{equation}

In the initial time interval $t_0$, the data amount carried by each vehicle is denoted as $A_{ini}$. If vehicle $k$ is not selected in time interval $t$, $A^k(t)$ does not change in time interval $t$.

\subsubsection{Transmission delay}
\
\newline
\indent
The transmission delay of vehicle $k$ in each time interval $t$ is the time it takes to send a packet to the RSU and can be calculated as
\begin{equation}
D^k(t)=\frac{D_a}{\rho^k(t)},
\label{eq6}
\end{equation}
where $\rho^k(t)$ is the transmission rate of vehicle $k$ in time interval $t$. The transmission rate model adopted is the wireless TCP/IP working in the steady-state \cite{39}, which can be expressed as
\begin{equation}
\rho^k(t) = \sigma^2c^k(t),
\label{eq7}
\end{equation}
where $\sigma^2$ is the variance of the signal strength estimation. Meanwhile, $c^k(t)$ is the correlation coefficient of vehicle $k$ in time interval $t$. The correlation coefficient is constructed based on the correlation model of shadow fading in mobile radio systems \cite{40}, which can be expressed as
\begin{equation}
c^k(t) = (\delta^k(t))^{(vt_s)/d_{k,r}(t)},
\label{eq8}
\end{equation}
where $v$ is the vehicle velocity, $d_{k,r}(t)$ is the distance from vehicle $k$ to RSU in time interval $t$, $t_s$ is the duration of a time interval and $\delta^k(t)$ is the correlation between two points separated by distance $d_{k,r}(t)$ in time interval $t$ (i.e., between vehicle $k$ and RSU) for the mobile radio system, which can be derived from the shadow fading model in \cite{40}.

Hence, given $D_a$, $\sigma^2$, $t_s$ and $v$, each vehicle can calculate the transmission delay in a time interval base on Eqs.~\eqref{eq6}-\eqref{eq8}.

\subsubsection{Packet collision probability}
\
\newline
\indent
The receiver (i.e., RSU $R_r$) may be prevented from receiving the packet correctly if both of the two cases occur, i.e., (1) the interference generated by interfering vehicle $i$, denoted by $v_i$, on $R_r$ is over the sensing power threshold $P_{sen}$; (2) transmitting vehicle $k$, denoted by $v_k$, and $v_i$ transmit simultaneously using the same resource. Thus, the packet collision probability caused by $v_i$ is composed of the probability under the two cases, denoted by $P_{col}^{i}(t)$, and can be calculated as
\begin{equation}
P_{col}^{i}(t)=P_{same}^{i}(t) \cdot P_{int}^{i}(t),
\label{eq9}
\end{equation}
where $P_{col}^{i}(t)$, $P_{same}^{i}(t)$ and $P_{int}^{i}(t)$ are the packet collision probability caused by $v_i$, the probability that $v_k$ and $v_i$ transmit simultaneously and the interference probability generated by $v_i$ over $P_{sen}$, respectively, in time interval $t$.

The packet collision probability of $v_k$ in time interval $t$, denoted as $P^k_{col}(t)$. Since the interfering vehicle $v_i$ may be any vehicle except $v_k$ in the communication range of the RSU, i.e., $(i \neq k)$, thus $P^k_{col}(t)$ can be calculated as
\begin{equation}
P^k_{col}(t)=1-\prod_{i(i \neq k)}\left[1-P_{col}^{i}(t)\right].
\label{eq10}
\end{equation}

Next, we will calculate $P_{int}^{i}(t)$ and $P_{same}^{i}(t)$, respectively.

\noindent \textbf{a.}  $P_{int}^{i}(t)$

To calculate $P_{int}^{i}(t)$, assume that the impact of the interference from $v_i$ on $R_r$ is equivalent to additional noise \cite{19}. Thus, the signal to interference plus noise ratio (SINR) in logarithmic mode in time interval $t$ can be calculated as \cite{19}
\begin{equation}
SINR(t)=P^{k}(t)-P^{i}(t)-N_{0},
\label{eq11}
\end{equation}
where $P^{k}(t)$ is the signal power received by $R_r$ from $v_k$ in time interval $t$, $P^{i}(t)$ is the signal power received by $R_r$ from $v_i$ in time interval $t$ and $N_0$ is the noise power. $P^k(t)$ and $P^i(t)$ can be calculated respectively as
\begin{equation}
P^{k}(t)=P-PL^k(t),
\label{eq12}
\end{equation}
\begin{equation}
P^{i}(t)=P-PL^i(t),
\label{eq13}
\end{equation}
where $P$ is transmission power, $PL^k(t)$ is the path loss power of $v_k$ in time interval $t$ and $PL^i(t)$ is the path loss power of $v_i$ in time interval $t$.

Next, we calculate $PL^k(t)$ and $PL^i(t)$. To calculate them, we introduce the urban micro (UMi) path loss model. In this model, the breakpoint distance of vehicle $x$ in time interval $t$ $d_b^x(t)$ ($x \in \left\{v_k, v_i\right\}$) is calculated as \cite{41}


\begin{equation}
d^x_{b}(t)=\frac{4 \cdot (h_V-h_{env}) \cdot (h_R-h_{env}) \cdot (f_c+f^x_d(t))}{c},
\label{eq14}
\end{equation}
where $h_V$ is the antenna height of vehicle $x$, $h_R$ is the antenna height of RSU $R_r$, $h_{env}$ is the average environmental height, $c$ is the velocity of light, $f_c$ is the carrier frequency, and $f^x_d(t)$ is the Doppler frequency of vehicle $x$ in time interval $t$, which can be calculated as\cite{42}

\begin{equation}
f^x_{d}(t)=\frac{v}{B} \cos \theta^x(t),
\label{eq15}
\end{equation}
where $B$ is the wavelength and $\cos\theta^x(t)$ is the cosine of the angle formed by the driving direction and the radio wave incident direction of vehicle $x$ in time interval $t$. They can be calculated respectively as
\begin{equation}
B=\frac{c}{f_c},
\label{eq16}
\end{equation}
\begin{equation}
\cos \theta^x(t)=\frac{d_{x,r}(t)}{m^x(t)},
\label{eq17}
\end{equation}
where $d_{x,r}(t)$ is the distance from the vehicle $x$ to $R_r$ in time interval $t$ and $m^x(t)$ is the distance traveled by the radio wave from vehicle $x$ in time interval $t$, which can be calculated based on Pythagorean theorem, i.e.,

\begin{equation}
m^x(t)=\sqrt{(h_R)^{2}+(d_{x,r}(t))^{2}}.
\label{eq18}
\end{equation}

According to the UMi path loss model, if $d_{x,r}(t)$ is smaller than $d^x_{b}(t)$, $PL^x(t)$ can be calculated as \cite{41}
\begin{equation}
PL^x(t)=22\log_{10}(d_{x,r}(t))+28+20\log_{10}(f_c+f^x_d(t)).
\label{eq19}
\end{equation}

According to the UMi path loss model, if $d_{x,r}(t)$ is lager than or equal to $d^x_{b}(t)$, $PL^x(t)$ can be calculated as \cite{41}
\begin{equation}
\begin{aligned}
&PL^x(t)=40\log_{10}(d_{x,r}(t))+7.8-18\log_{10}(h_V-h_{env})\\ &-18\log_{10}(h_E-h_{env})+2\log_{10}(f_c+fx_d(t)).
\end{aligned}
\label{eq20}
\end{equation}

Let $s$ be SINR, the probability density function of SINR when $P^{k}(t) \textgreater P_{sen}$ is denoted by $f_{SINR \mid P^{k}(t)>P_{sen}}(s)$, and it can be calculated by cross-correlating the probability density function of $P^k(t)$ and $P^i(t)$ according to probability theory \cite{43}. Let $BL(s)$ be the block error rate (BLER), which is a function of SINR for a given packet size \cite{44}. Since $BL(s) \cdot f_{SINR \mid P^{k}(t)>P_{sen}}(s)$ is the probability density function of SINR when $P^{k}(t)$ is higher than $P_{sen}$, thus the interference probability generated by $v_i$ over $P_{sen}$ in time interval $t$ can be calculated as

\begin{equation}
P_{int}^{i}(t)=\sum_{s=-\infty}^{+\infty} B L(s) \cdot f_{SINR \mid P^{k}(t)>P_{sen}}(s),
\label{eq21}
\end{equation}

\noindent \textbf{b.} $P_{same}^{i}(t)$

We assume that $N_T$ is the total number of resources in the selection window. Considering 1000 sub-frames per second, according to Fig.~\ref{fig2}, $N_T$ can be calculated as
\begin{equation}
N_T=\frac{1000 \cdot S}{f},
\label{eq22}
\end{equation}
where $S$ is the number of sub-channels per sub-frame and $f$ is the number of packets transmitted per second by each vehicle, i.e. the packet transmission frequency.

To calculate $P_{same}^{i}(t)$, we need to calculate the common resources in the candidate resource list $L_C$ for $v_k$ and $v_i$. We assume that each vehicle has the same sensing range, denoted by $d_{sr}$, and the number of vehicles in the sensing range of $v_k$ in time interval $t$ is $K_S(t)$. The number of vehicles in the common sensing range of $v_k$ and $v_i$ in time interval $t$ is $K_C(t)$. Vehicle $v_i$ may be driving in or out of the sensing range of vehicle $v_k$. If the vehicles do not occupy the same resources in the common sensing range of $v_k$ and $v_i$ in time interval $t$, the number of the resources occupied by the vehicles in the common sensing range $N_c(t)$ equals $K_C(t)$, i.e., $N_c(t) = K_C(t)$. However, if the vehicles occupy the same resource in the common sensing range of $v_k$ and $v_i$ in time interval $t$, $N_c(t) \leq K_C(t)$.

To calculate $N_c(t)$, the problem can be approximately modeled as the distribution of $K_C(t)$ distinguishable balls into $N_T$ distinguishable baskets. $N_c(t)$ can be deemed as the number of baskets containing at least one ball \cite{45}. For each basket, the probability that it contains at least one ball is $\left[1-\left(1-\frac{1}{N_T}\right)^{K_C(t)}\right]$.  Hence, $N_{c}(t)$ can be calculated as
\begin{equation}
N_{c}(t)=N_{T}\left[1-\left(1-\frac{1}{N_{T}}\right)^{K_C(t)}\right].
\label{eq23}
\end{equation}

According to the above approximate model, the number of the resources occupied by the vehicles in the sensing range of $v_k$ in time interval $t$ can be calculated as
\begin{equation}
N_{oa}(t)=N_{T}\left[1-\left(1-\frac{1}{N_{T}}\right)^{K_S(t)}\right].
\label{eq24}
\end{equation}

The number of the resources occupied by the vehicles in the sensing range of $v_k$ except the common sensing range of $v_k$ and $v_i$ in time interval $t$ can be calculated as
\begin{equation}
N_A(t)=N_{oa}(t)-N_{c}(t).
\label{eq25}
\end{equation}

Then, the number of the resources occupied by the vehicles that are not in the common sensing range of $v_k$ and $v_i$ in time interval $t$ can be calculated as
\begin{equation}
N_D(t)=N_{T}-N_{c}(t).
\label{eq26}
\end{equation}

According to the approximate model, the number of the common candidate resources of $v_k$ and $v_i$ in time interval $t$ is the probability that there is no ball in each basket, i.e., distribution of $N_A(t)$ distinguishable balls into $N_D(t)-N_A(t)$ distinguishable baskets, where $N_D(t)-N_A(t)$ is the number of the resources occupied by the vehicles in the sensing range of $v_i$ except the common sensing range of $v_k$ and $v_i$ in time interval $t$, thus it can be calculated as\cite{45}
\begin{equation}
N_{ccr}(t)=\left[N_D(t)-N_A(t)\right]\left[1-\frac{1}{N_D(t)}\right]^{N_A(t)}.
\label{eq27}
\end{equation}

According to the SPS scheme, each vehicle only selects one resource from its 20\% least interfered ones. Let $N_{lc}$ be the number of the resources in the list $L_C$ of candidate resources, thus $N_{lc}=0.2N_T$. Let $N_{ccr}(t)$ be the number of the common candidate resources of $v_k$ and $v_i$ in time interval $t$. For each one of the $N_{ccr}(t)$ resources, the probability $v_k$ and $v_i$ select the same resource is $\frac{N_{ccr}(t)}{N_{lc}^2}$. Based on the above analysis, the probability that $v_k$ and $v_i$ transmit simultaneously using the same resource in time interval $t$ can be calculated as\cite{45}
\begin{equation}
P_{same}^{i}(t)=\left\{\begin{array}{lc}
\frac{P_{RC=0} N_{ccr}(t)}{N_{lc}^2}, & d_{k, i}(t) \leq d_{sr}, \\
\frac{N_{ccr}(t)}{N_{lc}^2}, & d_{k, i}(t) \textgreater d_{sr},
\end{array}\right.
\label{eq28}
\end{equation}
where $P_{RC=0}$ is the probability of $RC$ decreased to zero, which can be calculated based on SPS scheme as

\begin{equation}
P_{RC=0}=\frac{1}{R_h-R_l}.
\label{eq29}
\end{equation}

Hence, given $d_{x,r}(t)$, $d_{k,i}(t)$, $P$, $h_V$, $h_R$, $h_{env}$, $f_c$, $c$, $P_{sen}$, $S$, $f$, $R_h$, $R_l$ and $d_{sr}$, vehicle $k$ can calculate the packet collision probability in time interval $t$, base on Eqs.~\eqref{eq9}-\eqref{eq29}.

\subsubsection{Survival ability}
\
\newline
\indent
The survival ability of vehicle $k$ in time interval $t$ is denoted as $S^k(t)$, which is the duration that vehicle $k$ stays in RSU's communication range in time interval $t$. It decreases by one unit time after every time interval, and can be calculated as

\begin{equation}
S^k(t) = S^k(t-1) - t_s.
\label{eq30}
\end{equation}

In the initial time interval $t_0$, the survival ability of vehicle $k$ can be calculated as
\begin{equation}
S_{ini}^k(t_0) = (E-d_{ini}^k)/v,
\label{eq31}
\end{equation}
where $E$ is the communication range of a RSU, $d_{ini}^k$ is the initial position of vehicle $k$.

Thus, given $E$, $d_{ini}^k$, $t_s$ and $v$, vehicle $k$ can calculate the survival ability in time interval $t$ base on Eqs.~\eqref{eq30} and ~\eqref{eq31}.

Based on Eqs.~\eqref{eq5}-\eqref{eq31}, the system status of each vehicle can be determined from the analysis above. Then, the vehicles will send their system statuses to the RSU, which will select the vehicles.

\subsection{New vehicle selection method}

The RSU selects vehicles for each time interval by following two steps. First, it estimates how many vehicles to select based on the current cache queue's backlog to prevent overflow. Second, it estimates which vehicles to select based on their priority, which is determined by their system status, to ensure more data amount participation in training. 


\begin{figure*}[htbp]
\centering
\includegraphics[width=12cm, scale=1.00]{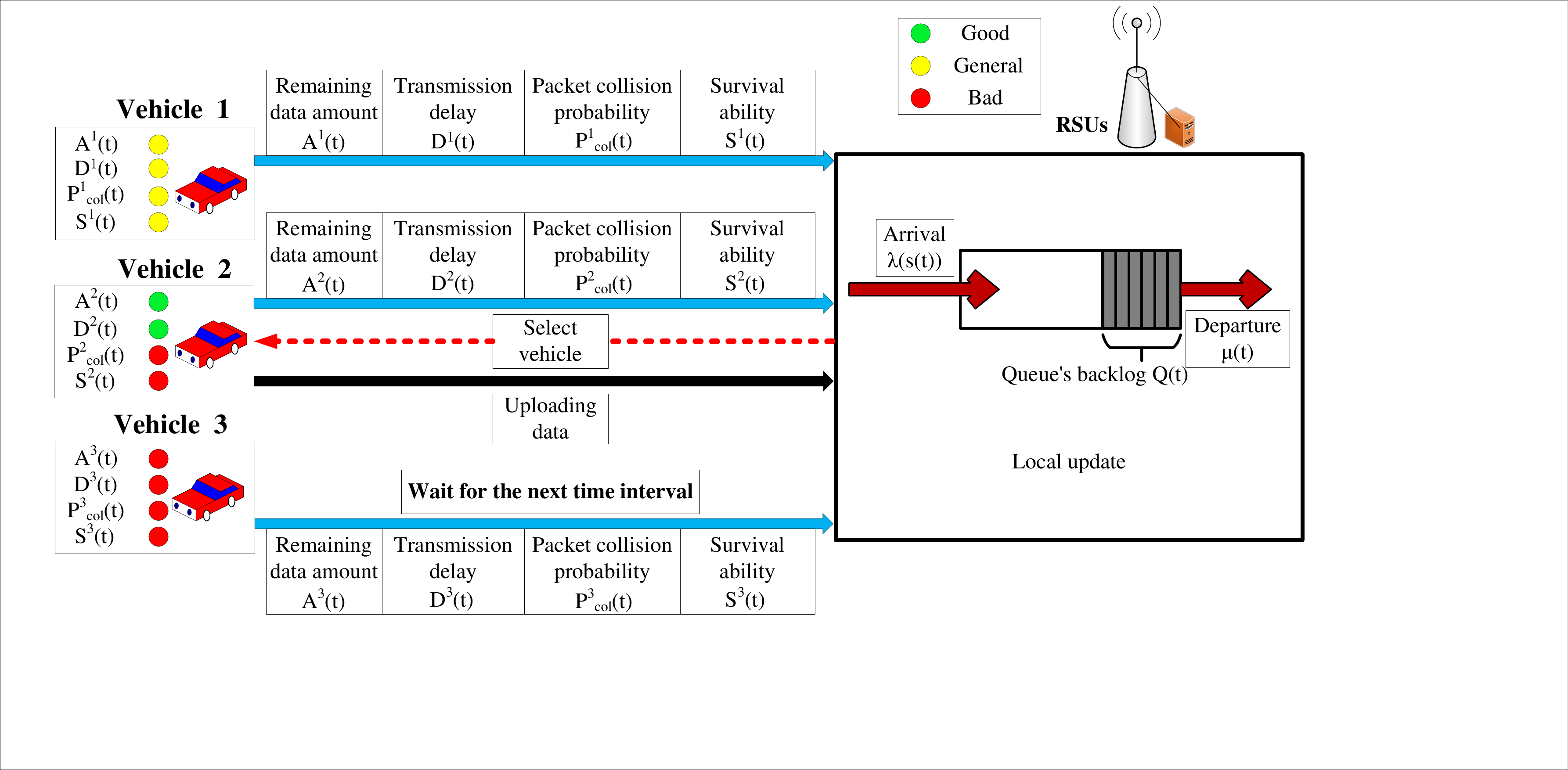}
\caption{RSU selection process for vehicles.}
\label{fig4}
\vspace{-0.7cm}
\end{figure*}

In time interval $t$, the RSU first receives $A^k(t)$, $D^k(t)$, $P^k_{col}(t)$ and $S^{k}(t)$ from each vehicle $k$, and then iteratively determines $s^*(t)$. As explained in Section~\ref{Problem Formulation}, the goal is to find $s^*(t)$. In each iteration, if the queue's backlog is not larger than the $Q_{max}$ after data reach the queue, i.e., $0\leq Q(t-1)+\lambda(s(t)) \leq Q_{max}$, the value of optimization objective, which is denoted as $Q$, is calculated base on Eq.~\eqref{eq3}, then $s(t)$ is increased by one. Otherwise, if $Q(t-1)+\lambda(s(t))>Q_{max}$, $Q$ is assigned as $Q^*$ and $s(t)-1$ is assigned as $s^*(t)$. The above procedures will be repeated until $s^*(t)$ can be determined.

Next, the RSU decides the vehicles that should be selected. In particular, the RSU first calculates the priority of each vehicle $k$ according to its system status. The priority is defined by the following cases. Some vehicles with a large remaining data amount may not have the opportunity to upload all their data if the vehicles with a small remaining data amount are selected. In addition, the vehicle will upload more data when the transmission delay is low, and the vehicle with a lower packet collision probability is selected to improve successful packet transmission. Moreover, the vehicles with small survival ability are preferred to be selected because they will soon drive out of the RSU's communication range. Thus, if $P_{col}^{k}(t)$ and $S^{k}(t)$ are not 0, the priority of vehicle $k$ is calculated as
\begin{equation}
p^k(t) = \frac{A^k(t)}{D^k(t) \cdot P^k_{col}(t) \cdot S^{k}(t)},
\label{eq32}
\end{equation}
or else, $p^k(t)$ is 0. 

Afterwards, $s^*(t)$ vehicles with the largest priorities can be used for data uploading. During data uploading, if a collision of a packet has occurred, the packet is discarded. Meanwhile, $s^*(t)$ is reduced by one. 
After the data of the selected vehicles are received, the RSU adopts the convolutional neural network to train a local model and obtain the training accuracy. Finally, the queue's backlog is updated by RSU based on Eq.~\eqref{eq1}, and the above steps are repeated in the next time interval.

For ease of understanding, Fig.~\ref{fig4} illustrates the process of vehicle selection at the RSU. In Fig.~\ref{fig4}, vehicle 1 and 2 have been selected owing to their desirable system status, while vehicle 3 is not selected because of its bad system status and need to wait for the next time interval.

\section{Simulation Results and Analysis}
\label{Simulation Results and Analysis}

In this section, we perform extensive simulation experiments and numerical evaluations to verify the performance of the vehicle selection method by comparing key performance indicators. The simulation is based on Python 3.6 and MATLAB 2019b. Python 3.6 is used to implement the FEEL system for selecting vehicle, while MATLAB 2019b is used to implement the C-V2X Mode 4 communication process. The MATLAB code is realized based on the code in related work \cite{19}, which has been widely adopted by researchers. According to \cite{19}, the results conducted by MATLAB have been validated by Veins, which is a tool to simulate the real network protocol and environment. Hence, our results can reflect the critical performance metrics of the proposed algorithm. The model training is based on the publicly available MNIST dataset. The simulation scenario is a one-way road covered by multiple RSUs. It is assumed that each RSU in the vehicle network is equipped with a cache queue and covers $K$ vehicles, and $K$ vehicles driving at the same velocity, where each vehicle carries 1500 bytes of data. In each time interval (i.e., 100 ms), a selected vehicle uploads 10 bytes of data to the RSU. According to LTE standard, the sensing power threshold is set as -90.4 dBm, and when $f$=10, $\Gamma$=100, $R_h$=15, $R_l$=5, when $f$=20, $\Gamma$=50, $R_h$=30, $R_l$=10, and when $f$=50, $\Gamma$=20, $R_h$=75, $R_l$=25 \cite{37}. Table~\ref{tab1} shows the parameters used in the simulation.

\begin{figure}[htbp]
    \centering
    \includegraphics[width=0.5\textwidth]{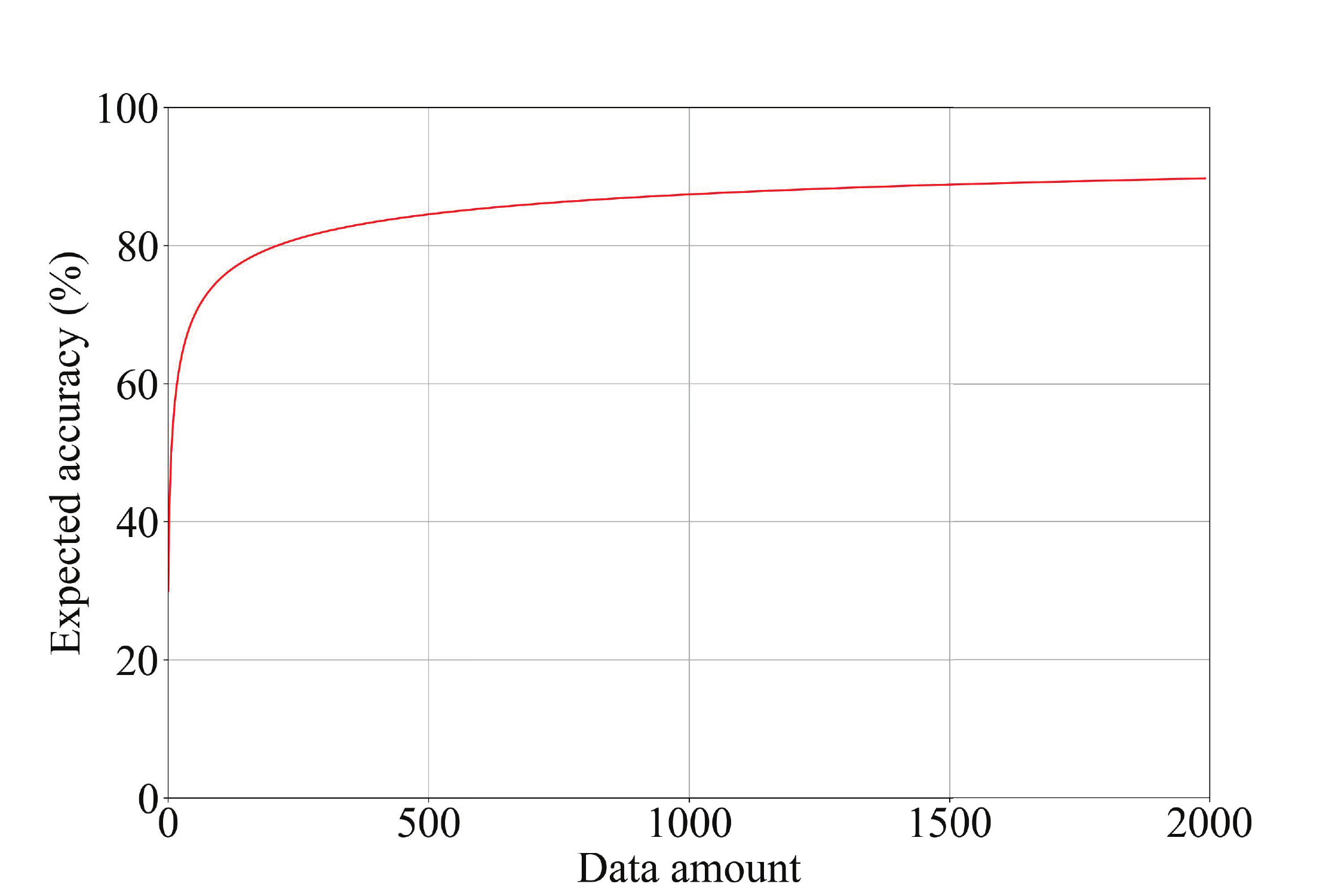}
    \caption{Expected accuracy.}
	\label{fig5}
\vspace{-0.7cm}
\end{figure}

Fig.~\ref{fig5} illustrates the learning curve as the data amount increases, and the learning curve is the expected accuracy of the utility function $U(s(t))$ \cite{46}. When the training data amount is $x$, the expected accuracy can be calculated as


\begin{table}
\caption{Values of the parameters in our simulation.}
\label{tab1}
\footnotesize
\centering
\begin{tabular}{|c|c|c|c|}
  \hline
  \textbf{Parameter} &\textbf{Value} &\textbf{Parameter} &\textbf{Value}\\
  \hline
  {$K$} &50,100,150,200 &{$Q_{max}$} &2000 bytes\\
  \hline
  {$E$} &1000 m &{$R$} &500 m\\
  \hline
  {$\gamma$} &$10^{9}$ &{$f$} &10, 20, 50 Hz \\
  \hline
  {$\Gamma$} &20, 50, 100 ms  &{$R_h$} &15, 30, 75 \\
  \hline
  {$R_l$} &5, 10, 25 &{$\sigma^2$} &7.5 dB\\
  \hline
  {$A_{ini}$} &1500 bytes &{$D_a$} &10 bytes\\
  \hline
  {$v$} &36 km/h  &{$P$} &23 dBm\\
  \hline
  {$S$} &2, 4 &{$f_c$} &$2.5 \times 10^9$ Hz \\
  \hline
  {$h_V$} &1.5 m &{$h_R$} &10 m \\
  \hline
  {$h_{env}$} &0 m &{$\alpha$} &$5.6 \times 10^6$ \\
  \hline
  {$d_{sr}$} &500 m &{$l_{rate}$} &1 \\
  \hline
  {$d_{rate}$} &-0.3 &{$P_{sen}$} &-90.4 dB \\
  \hline
\end{tabular}
\end{table}

\begin{equation}
E_{acc}(x) = 1-l_{rate} \cdot x^{d_{rate}},
\label{eq33}
\end{equation}
where $l_{rate}$ is the learning rate and $d_{rate}$ is the decay rate.

In order to show a generic result, we select the average value over 6 times of experiments for all the below results.

\begin{figure}[htbp]
\vspace{-0.5cm}
    \centering
    \includegraphics[width=0.5\textwidth]{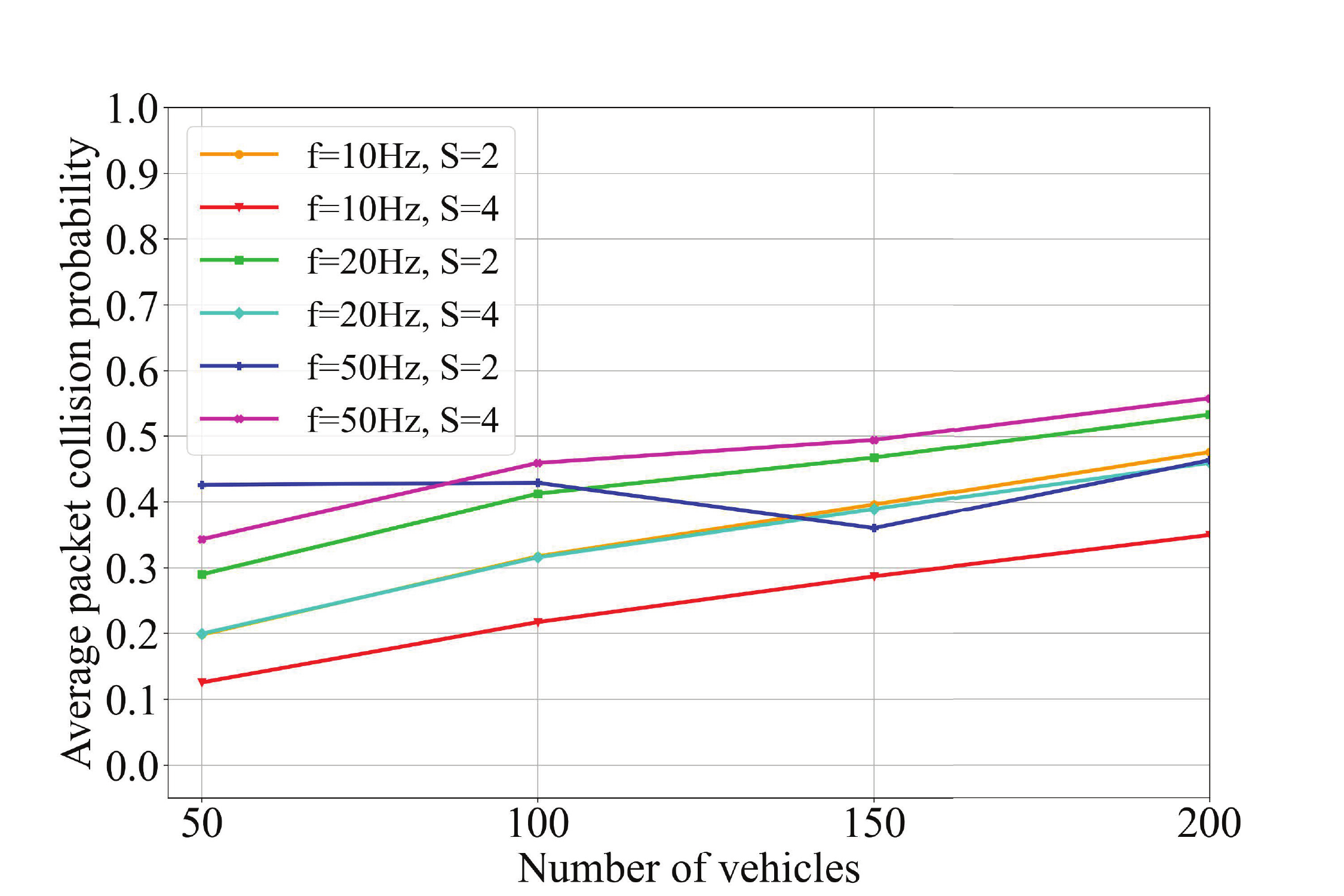}
    \caption{The variation of the average packet collision probability with the number of vehicles (transmitted power $P$=23 dBm).}
	\label{fig6}
\vspace{-0.7cm}
\end{figure}

Fig.~\ref{fig6} shows the average packet collision probability with $K$ vehicles under different packet transmission frequencies $f$ and numbers of sub-channels $S$. Note that vehicles are in the initial position and the transmission power $P$ is 23 dBm. It is seen that the average packet collision probability increases as the number of vehicles increases for all cases except that when $f$=50 Hz and $S$=2. The average packet collision probability is higher when the packet transmission frequency is higher and the number of sub-channels is fewer. This is because the packet collision probability is composed of two parts: the interference probability generated by the interfering vehicles is over the sensing power threshold and the probability that two vehicles transmit simultaneously using the same resource. Given a larger number of vehicles, a higher packet transmission frequency, and a fewer number of sub-channels, the above two probabilities will be increased accordingly, and thus contribute to the increasing packet collision probability. However, one can see that when $f$=50 Hz and $S$=2, the average packet collision probability decreases at first and then starts to grow as the number of vehicles increases. This is because given $f$=50Hz and $S$=2, the total number of resources in the selection window $N_T$ calculated in Eq.~\eqref{eq22} is too small, and thus the sensing power threshold of Step 2 in the SPS process becomes larger after some iterations. In this case, the interference probability is over the sensing power threshold will decrease incurring a low average packet collision probability. In addition, because $N_T$ is too small, the number of the common candidate resources $N_{ccr}$ decreases in a crowded scenario. This affects the probability that two vehicles transmit simultaneously using the same resource, and thus the average packet collision probability decreases.

\begin{figure}[htbp]
    \centering
    \includegraphics[width=0.5\textwidth]{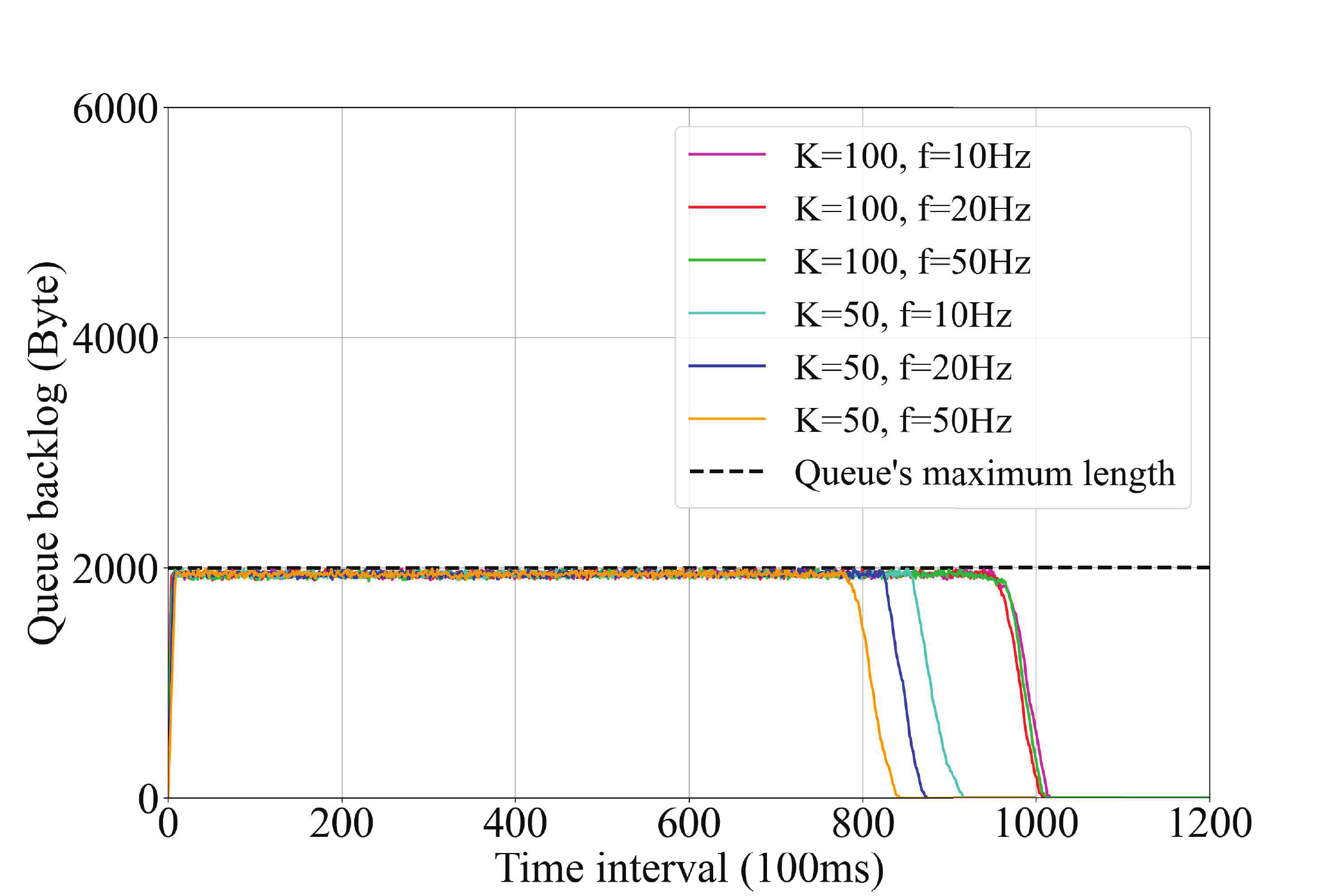}
    \caption{Queue's backlog comparison for $S$=2.}
	\label{fig7}
\end{figure}

Fig.~\ref{fig7} compares the queue's backlog with different numbers of vehicles $K$ and packet transmission frequencies $f$ when the number of sub-channels $S$=2. The black dotted line is the $Q_{max}$. The observation shows that the queue's backlog keeps stability for more time intervals under $K$=100 than that under $K$=50, and the queue's backlog keeps stability for more time intervals when $f$ is smaller for all cases except when $f$=50 Hz and $S$=2. This is because when $K$=100, there are more vehicles being selected and more data involved in training. Also, the packet collision probability is lower when $f$ is smaller. Therefore, more vehicles can be selected to upload data and thus keeps more time intervals. However, in a crowded scenario, i.e., when the number of vehicles $K$=100 ($f$=50 Hz, $S$=2), the total number of resources in the selection window $N_T$ is too small, which results in a larger sensing power threshold and a reduced interference probability is over the sensing power threshold, thus incurring a low packet collision probability, and the queue's backlog keeps stability for more time intervals than that under $f$=20 Hz.

\begin{figure}[htbp]
    \centering
    \includegraphics[width=0.5\textwidth]{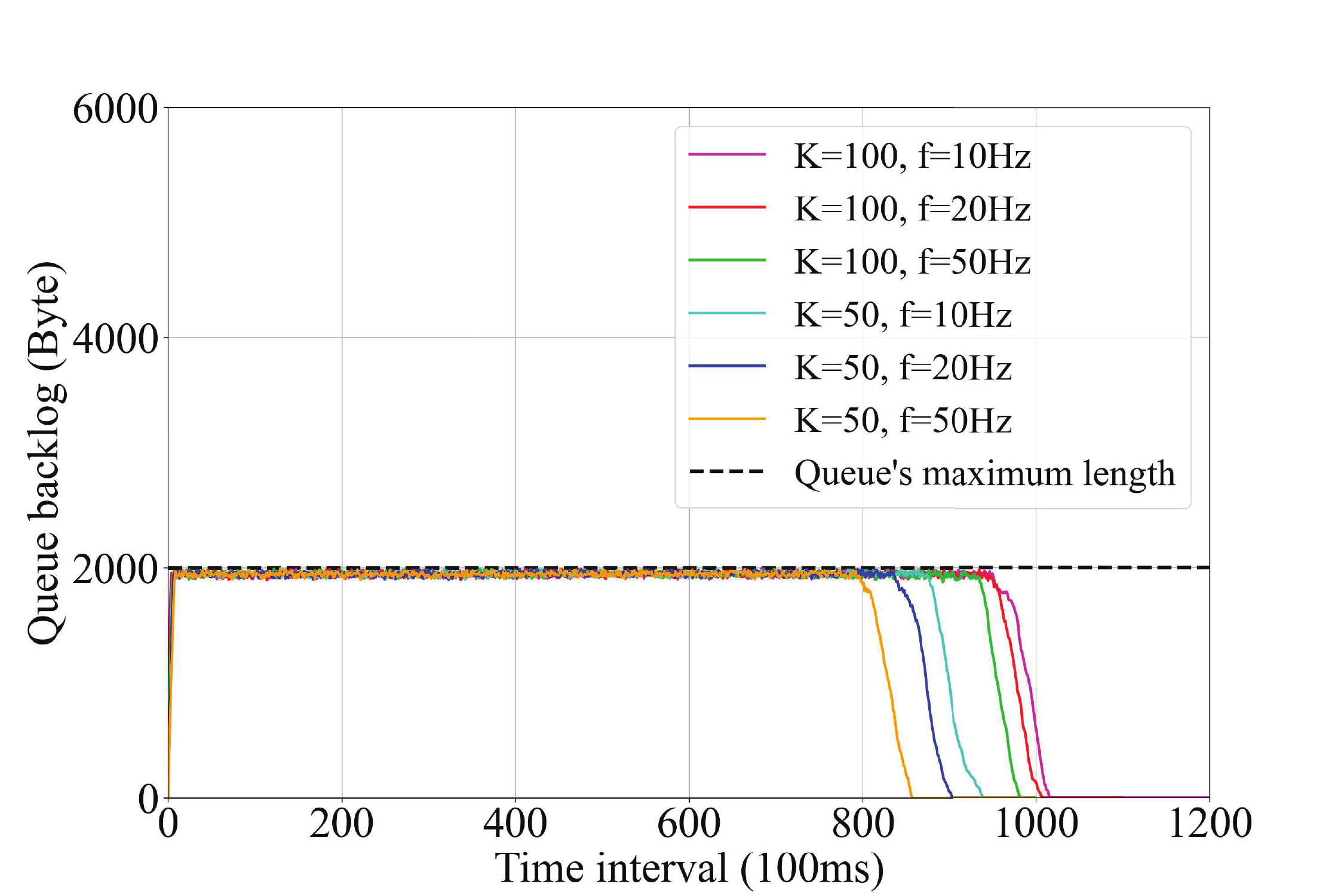}
    \caption{Queue's backlog comparison for $S$=4.}
	\label{fig8}
\end{figure}

Fig.~\ref{fig8} compares the queue's backlog with different numbers of vehicles $K$ and packet transmission frequencies $f$ when the number of sub-channels $S$=4. Similar to Fig.~\ref{fig7}, the queue's backlog keeps stability for more time intervals when $K$=100 and $f$ are smaller. The reason is also similar to the case when $S$=2. However, when $S$=4, the queue's backlog keeps stability for more time intervals than that when $S$=2. It is because that when $S$=4, there are more total resources, which results in lower packet collision probability, and thus more vehicles can successfully upload data.

Based on the analysis of the above experimental results, we choose $f$=10Hz, $S$=4 for the following simulation experiments to make the queue's backlog keep stability for more time intervals.

\begin{figure}[htbp]
    \centering
    \includegraphics[width=0.5\textwidth]{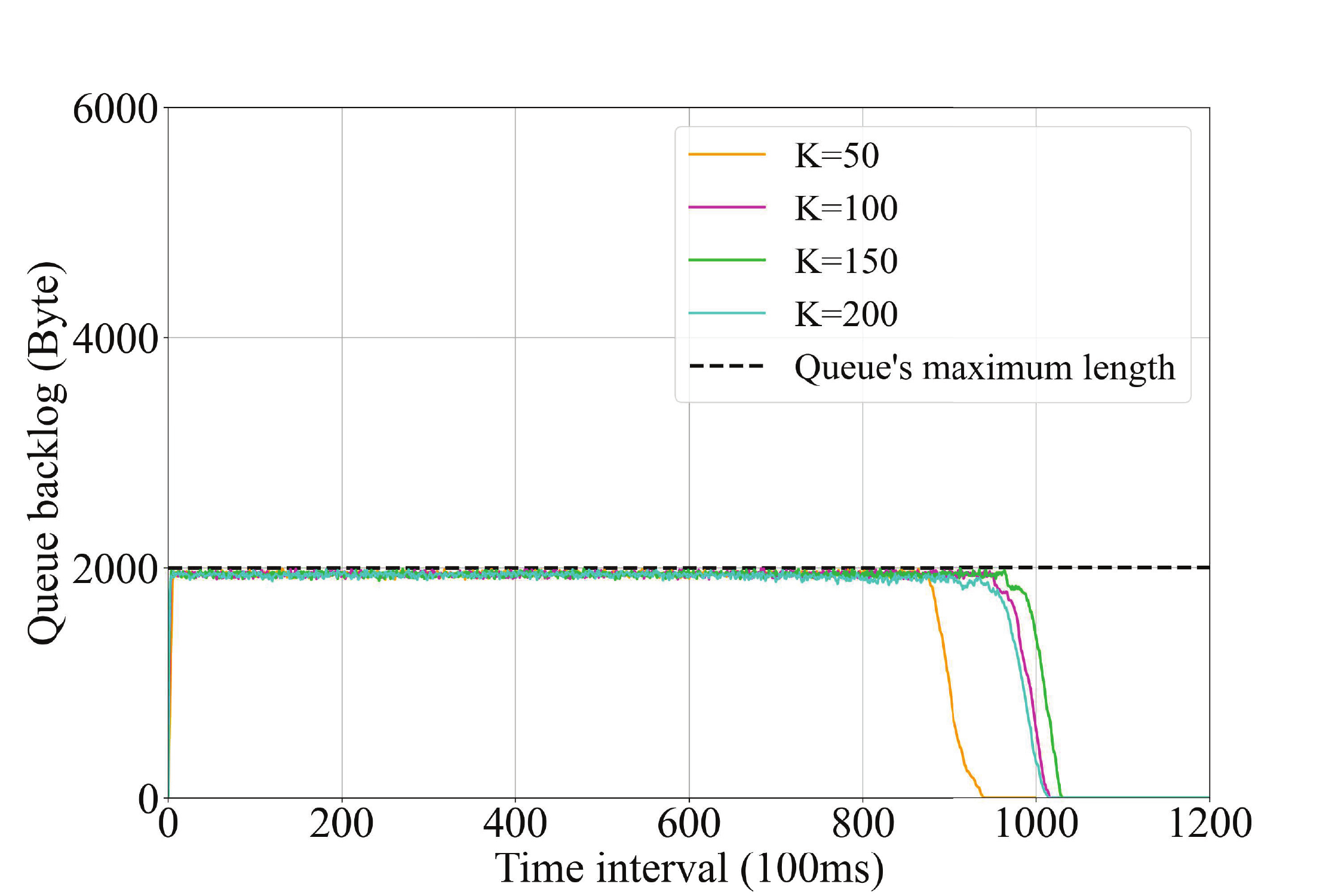}
    \caption{Comparison of the queue's backlog with different numbers of vehicles ($S$=4 and $f$=10 Hz).}
	\label{fig9}
\end{figure}

Fig.~\ref{fig9} compares the queue's backlog with different numbers of vehicles $K$ when $S$=4 and $f$=10 Hz. The black dotted line is the $Q_{max}$. The observation shows that the queue's backlog keeps stability for less time intervals under $K$=50, while the other cases experience around 1020 time intervals. This is because the larger $K$ can lead to a larger total data amount in the communication range of the RSU, which may experience more time intervals. However, given a larger $K$, the packet collision probability would increase, and thus a large data amount cannot be successfully transmitted. Hence, except for $K$=50, the time intervals experienced in other cases are basically similar. Therefore, we choose the number of vehicles $K$=100 to complete the subsequent simulation experiments.

In addition, we have compared our proposed vehicle selection method with four other baseline selection methods to evaluate its performance\cite{17, 21}, i.e.,

\noindent \textbf{Maximum selection method}: The RSU receives all vehicles' data in each time interval\cite{47, 48}.

\noindent \textbf{Static selection method}: The RSU selects the same number of vehicles at random in each time interval. In our simulation, each time interval has 4 vehicles selected randomly for data transmission\cite{7}.

\noindent \textbf{Random selection method}: The method for estimating $s(t)$ is identical to our proposed method; but, the selection of vehicles is random and does not consider the system status of different vehicles\cite{7, 48, 49}.

\noindent \textbf{Position-based selection method}: The RSU selects 10 vehicles with the closest distance to the RSU \cite{21}.

\begin{figure}[htbp]
    \centering
    \includegraphics[width=0.5\textwidth]{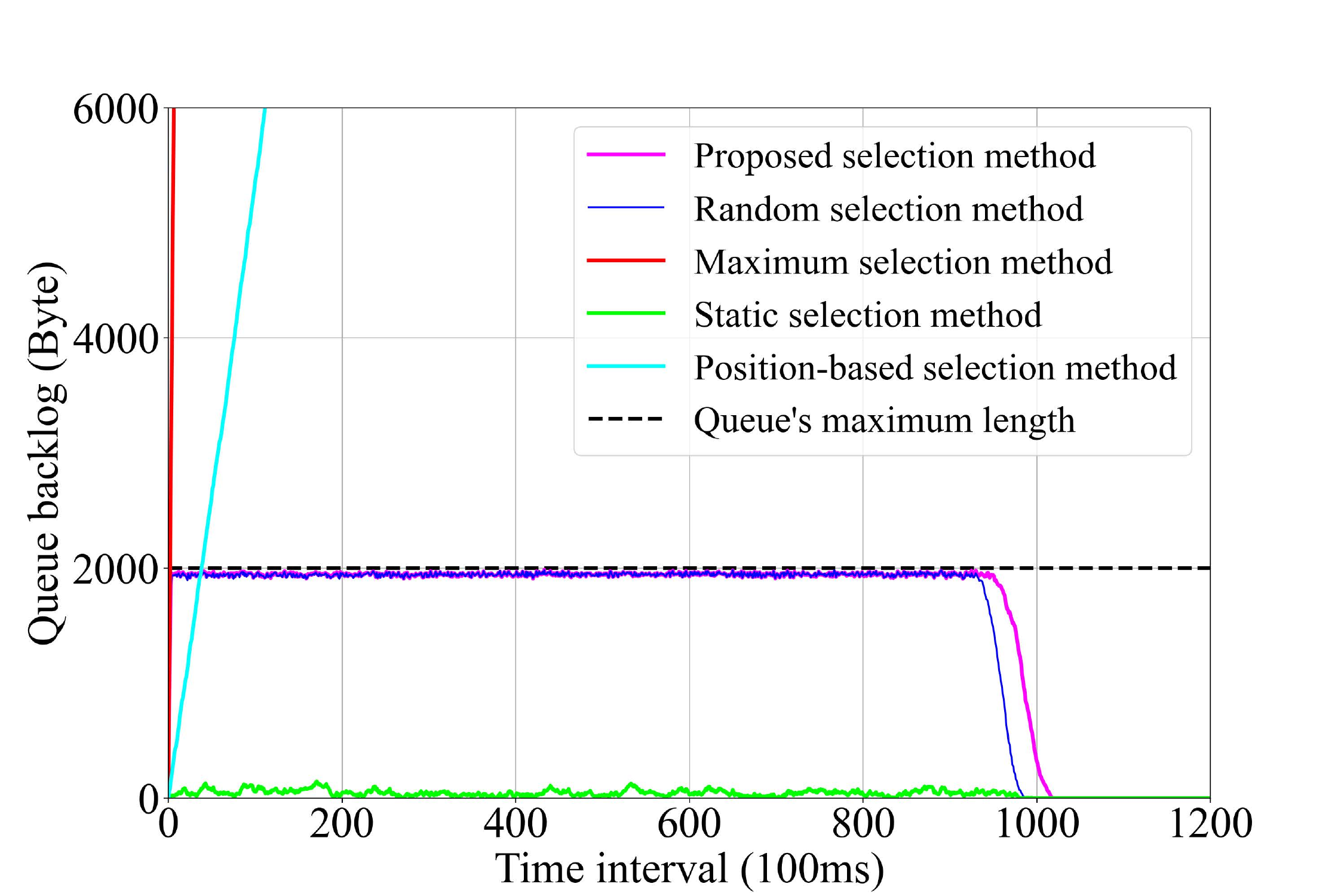}
    \caption{Comparison of the queue's backlog with different selection methods.}
	\label{fig10}
\end{figure}

Fig.~\ref{fig10} compares the queue's backlog of our proposed selection method with the four other selection methods. The black dotted line is the $Q_{max}$. The observation shows that when the maximum selection method is used, the queue's backlog quickly increases and surpasses $Q_{max}$. This verifies that the cache queue can become overloaded and unstable if too many vehicles upload data. The queue's backlog is much lower than $Q_{max}$ when the static selection method is used. This is because the static selection method selects a few vehicles to upload data in each time interval, resulting in very low utilization of the queue. When the position-based selection method is adopted, the queue backlog increases rapidly and exceeds the queue's maximum length after the 50th time interval, making the cache queue very unstable. This is because the position-based selection method selects too many vehicles with the closest distance to the RSU to upload data. When comparing our proposed method with the random selection method, the queue's backlog remains almost constant at $Q_{max}$ before the 950th time interval. After this, the queue's backlog under the random selection method slowly drops, but the queue's backlog under our proposed selection method still maintains $Q_{max}$ and drops after the 980th time interval, which shows the advantage of our proposed method. This is because both the random selection method and our proposed method use the same method to determine $s^*(t)$, so the queue's backlog is maintained at $Q_{max}$ before the 950th time interval. However, the random selection method does not consider the system status of vehicles in random vehicle selection. As a result, some vehicles became unavailable earlier than our proposed selection method due to insufficient remaining data or insufficient survival ability, resulting in no data available for the RSUs.

\begin{figure}[htbp]
\vspace{-0.5cm}
    \centering
    \includegraphics[width=0.5\textwidth]{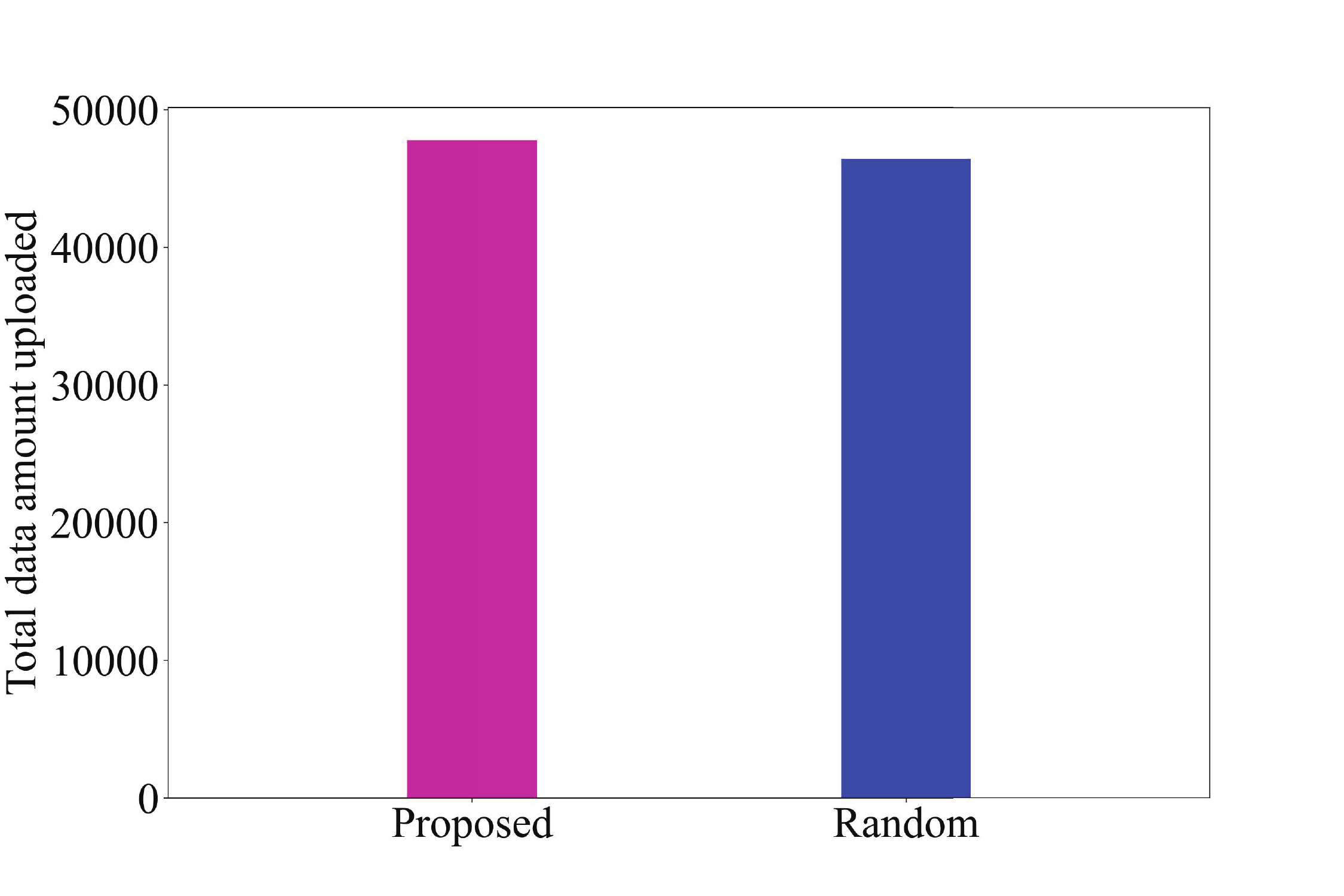}
    \caption{Comparison of the total data amount uploaded.}
	\label{fig11}
\end{figure}

Fig.~\ref{fig11} compares the total data amount uploaded by the selected vehicles under our proposed and random selection methods. The total data amount uploaded in our proposed selection method is 47760 bytes, while that in the random selection method is 46428 bytes. This is because our proposed selection method takes into account the heterogeneous system characteristics of vehicles, and more data are used in the training than the random selection method.

\begin{figure}[htbp]
    \centering
    \includegraphics[width=0.5\textwidth]{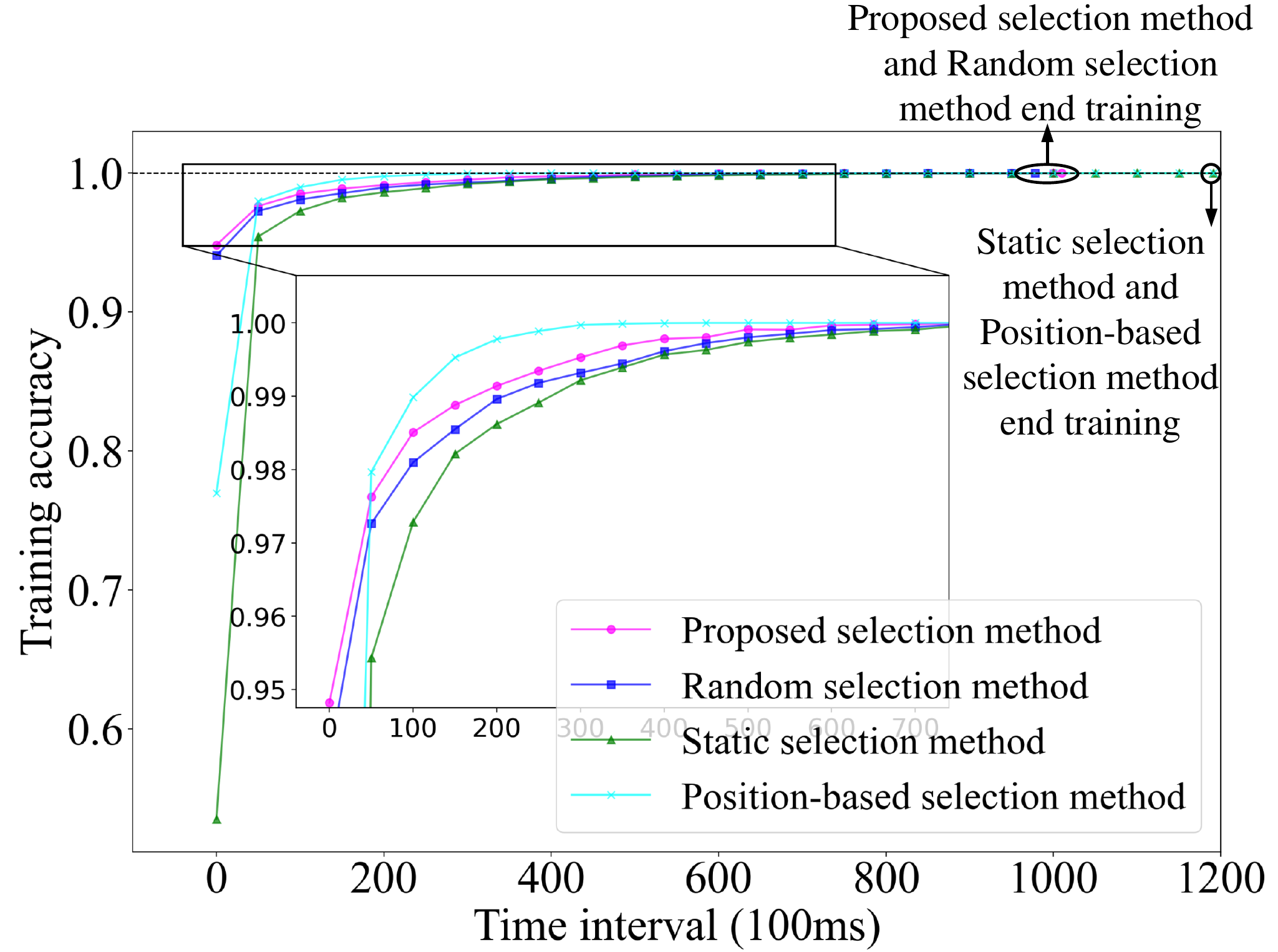}
    \caption{Comparison of training accuracy with different selection methods.}
	\label{fig12}
\end{figure}

Fig.~\ref{fig12} compares the training accuracies with different selection methods. We can see that when the static selection method is adopted, the training accuracy is lower than that with other methods. This verifies that if few vehicles upload data, the cache queue would have insufficient data for training and the accuracy of the model will suffer. The observation shows that the training accuracy of our proposed selection method is higher than that of the random selection method and the static selection method from the beginning. From the 600th time interval, the accuracy curves of the three selection methods gradually overlap. This is because our proposed selection method takes into account the system status of vehicles, which makes more vehicles participate in the federated training, and thus improves the training accuracy as compared to two other selection methods. In the 600th time interval, because the training contains enough data, all three selection methods start to converge. In addition, our proposed selection method does not end the training until the 1100th time interval because it considers the heterogeneous system characteristics of the vehicles, while the random selection method ends the training at the 970th time interval. Moreover, the static selection method selects a few vehicles to upload data in each time interval. Therefore, after the other two selection methods have finished their training, the static selection method can still select vehicles because there are still candidate vehicles in the communication range of the RSU. Therefore, the static selection method can be trained for a longer time until the maximum time interval is reached, but it can only select a few vehicles to upload data in each time interval. Although the position-based selection method has higher training accuracy and converges faster than other methods, its queue backlog exceeds the queue's maximum length early, making the cache queue unstable. This is because the position-based selection method selects too many vehicles with the closest distance to the RSU in each time interval.

\begin{figure}[htbp]
    \centering
    \includegraphics[width=0.5\textwidth]{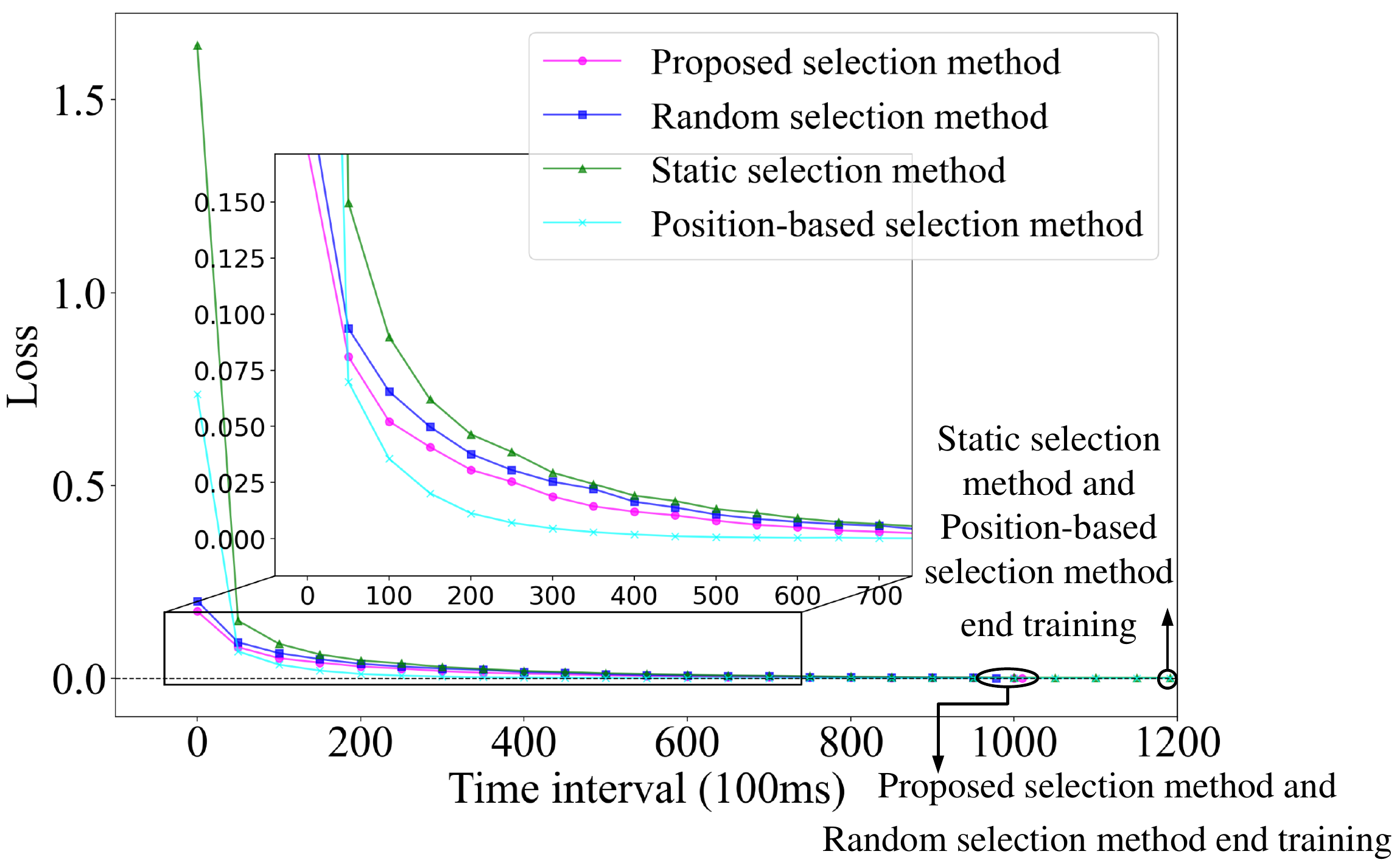}
    \caption{Comparison of loss with different selection methods.}
	\label{fig13}
\end{figure}

Fig.~\ref{fig13} compares the loss with different selection methods. The observation shows that the training loss of our proposed selection method is lower than the random selection method and static selection method in the beginning. The loss curves of the three selection methods gradually overlap and converge until the 600th time interval. The random selection method ends training at the 970th time interval, while our proposed selection method does not end training until the 1100th time interval and the static selection method can proceed until the last time interval. Moreover, the position-based selection method has lower loss and faster convergence than other methods. The reason is similar to the above training accuracy comparison figure.

\section{Conclusion}
\label{Conclusion}
In this paper, we proposed a vehicle selection method to maximize the accuracy of the model and maintain the cache queue stability for the FEEL systems. We first considered the random departure of data and determine the optimal number of selected vehicles to ensure the stability of the cache queue according to Lyapunov's control theorem. Then, we jointly considered the vehicles with different system status and proposed the vehicle selection method to maximize the accuracy of the model. Our proposed method outperforms other baseline selection methods according to extensive simulation experiments. The conclusions of this paper are listed as follows.

(1) For the C-V2X mode 4 based FEEL systems, the queue's backlog keeps stability for more time intervals under the smaller packet transmission frequency and the larger number of sub-channels. When the numbers of vehicles are 100, 150 and 200, the queue's backlog keeps stability for almost the same time intervals and the kept time intervals are more than that when the number of vehicles is 50.

(2) Our proposed method experiences more time intervals and uploads more data for model training than other baseline selection methods.

(3) Our proposed method converges faster in model training in terms of accuracy and loss than other baseline selection methods.








\vfill

\end{document}